\documentclass{article}

\usepackage[final]{neurips_2020_tda}

\usepackage{natbib}
\usepackage[utf8]{inputenc} 
\usepackage[T1]{fontenc}    
\usepackage{amsfonts}       
\usepackage{booktabs}       
\usepackage{hyperref}       
\usepackage{url}            
\usepackage{microtype}      
\usepackage{nicefrac}       
\usepackage{paralist}       
\usepackage{siunitx}        
\usepackage{dsfont}
\usepackage{amsmath}
\usepackage{graphicx}
\graphicspath{ {/} }

\title{
  Using topological autoencoders as a filtering function for global and local topology
}

\author{Filip Cornell \\ KTH Royal Institute of Technology \\ Kistagången 16, 164 40 Kista, Sweden \\ \href{mailto:fcornell@kth.se}{fcornell@kth.se}}

\begin{document}

\maketitle

\begin{abstract}
Choosing a suitable filtering function for the Mapper algorithm can be difficult due to its arbitrariness and domain-specific requirements. Finding a general filtering function that can be applied across domains is therefore of interest, since it would improve the representation of manifolds in higher dimensions. In this extended abstract, we propose that topological autoencoders is a suitable candidate for this and report initial results strengthening this hypothesis for one set of high-dimensional manifolds. The results indicate a potential for an easier choice of filtering function when using the Mapper algorithm, allowing for a more general and descriptive representation of high-dimensional data.
\end{abstract}

\section{Introduction}

Topological Data Analysis (TDA) has recently gained a great amount of popularity in the machine learning community. Apart from Persistent Homology, one of the more prominent methods has been the Mapper, originally proposed by \cite{singh2007topological} due to its ability to provide a visualization of high-dimensional data. In short, the Mapper algorithm, as explained in the original paper (see \cite{singh2007topological}) consists of a few steps. Given a point cloud $X \in \mathbb{R}^d$, one uses a filtration function $f : \mathbb{R}^d \rightarrow \mathbb{R}^m$, where $m < d$. This function, as stated by \cite{singh2007topological}, \textit{"can be a function which reflects geometric properties of the data"}. Once reduced using the filtrations, the points are clustered using a clustering method of choice and mapped into overlapping bins forming vertices connected if they have common points, producing a graph that serves as an approximation of the Reeb graph (see \cite{munch2015convergence}). 

One of the difficulties with the Mapper is however the choice of parameters such as deciding on the overlap between the bins and design of the filtering function $f$. These choices determine the construction of the resulting graph, and which filtering function that best captures the topology is not always entirely clear. While the choice of function allows for certain freedom, it might also pose as a barrier of entry for the regular practitioner and a difficulty in generating useful representations when using the Mapper algorithm. The optimal is of course to reduce the dimensionality in a way such that the local and the global topology of the data is preserved, while the summarization of the data lose the minimum amount of information. However, as \cite{kumari2020shapevis} points out, different filtering functions capture different things. \cite{carriere2018statistical} mention that the two most common choices are the eccentricity, i.e., the maximum distance to any other point in the point cloud, and the eigenfunctions of the covariance matrix as used in PCA. Other methods such as SVD (\cite{lum_singh_lehman_ishkanov_vejdemo-johansson_alagappan_carlsson_carlsson_2013}) as well as the kernel density estimation (\cite{singh2007topological}) have been extensively used. These do however all come with their drawbacks. SVD and PCA do not capture non-linear relations very well and might miss important non-linear properties of the data, although they preserve distance. Non-linear techniques such as t-SNE (\cite{maaten2008visualizing}) and IsoMAP (\cite{tenenbaum2000global}) are better at preserving local distances but lose the global perspective as pointed out by \cite{kumari2020shapevis}. The kernel density estimator relies on a bandwidth $\epsilon$, set by the user, introducing another parameter to tune. The choice of filtering function is therefore a choice for the user where trade-offs have to be made. Finding a filtering function without the aforementioned drawbacks and which could simultaneously capture both the global and local topologies would therefore be highly beneficial, since it would produce Mapper graphs representing the shape of the data more accurately. 

\section{Topological Autoencoders as a filtering function}

We propose the idea that a candidate for a filtering function preserving global and local topology is the topological autoencoder (TAE), introduced by \cite{moor2019topological}. These are autoencoders (AEs) regularized with a Persistent Homology loss to preserve the topology. As seen in their paper (see \cite{moor2019topological}), the TAEs manage to preserve the shape and form of the data in lower dimensions to a certain extent, and manages to capture the global as well as local topology apart from other methods. As they also note, the TAE can learn multiple manifolds, which is a difficult task for many methods. When this topology is preserved for a distance metric of choice, we can derive a compact representation in lower dimensions that better represents the data's shape. TAEs do of course come with a longer training time, a risk of overfitting, neural architecture choices and a need for larger datasets. It does however provide a bridge between topology and deep learning. 

\section{Initial experiments}

To evaluate whether TAEs result in more representative graphs than other filtering functions, we are looking at how the Mapper maps data belonging to different manifolds into vertices with different filtering functions. As a start, we use the synthetic data constructed by \cite{moor2019topological}, consisting of ten 100-dimensional hyperspheres inside one larger 100-dimensional hypersphere in 101 dimensions. We scale the data and use DBSCAN (see \cite{ram2010density}) as clustering method and set its $\varepsilon$ to 4 for all comparisons and the euclidean distance as the distance metric. For the TAE, we reproduce the experiments from \cite{moor2019topological}, with the exact same parameter choices as in their best runs. We use the same training data to train the representations and construct the Mapper graph, to then map the test data into the Mapper graph. 

Our initial experiments\footnote{Code available at \url{https://github.com/Filco306/Topo-AE-Mapper}.} aim for a high separation of the spheres into different vertices as well as separate simplicial complexes for them, working under the presumption that vertices should contain points generated from only one sphere. We use the metric $m_{k, o, i}$ as seen in equation \ref{eq:eval-result}, where $V_{k, o, i}$ is the set of vertices in the mapper graph $\mathcal{G}_{k, o, i} := (V_{k, o, i}, E_{k, o, i})$ constructed using filtering function $k$, overlap $o$ and number of intervals $i$. $\mathcal{C} = \{c_1, ..., c_n\}$ is the set of different manifolds in the dataset to which each point can belong. In our initial experiments, $k \in \{\mbox{T-SNE}, \mbox{TAE}, \mbox{Kernel Density}, \mbox{SVD}, \mbox{PCA}, \mbox{Eccentricity}\}$ and $\mathcal{C} = \{c_1,..,c_{11}\}$, one for each hypersphere. For the trainable filtering functions, we train our filtering functions on a training set of data and visualize and build the graph on a test set. However, for TSNE, which is not made for predicting mapping points to its latent space, we use as many points (9000) to train, but only produce a graph on the test set points for a fair comparison.  We use a covering with overlappings rangings from 0.025 to 0.4 with a 0.025 step and an interval cardinality ranging between 5 and 45 with step of 5 and report the best and worst results in table \ref{table:res}. In other words, $o \in \mathcal{O} := \{0.025, 0.05, ..., 0.4\}$ and $i \in \mathcal{I} := \{5, 10, ..., 45\}$. Running over many choices can bring insights into a filtering function's robustness to choices of the aforementioned parameters. 

\begin{equation} \label{eq:eval-result}
m_{k, o, i} = \frac{1}{|V_{k, o, i}|}\sum_{v \in V_{k, o, i}} \sum_{c \in \mathcal{C}} \mathds{1}(c \in v), \forall o \in \mathcal{O}, i \in \mathcal{I}
\end{equation}

The intuition behind our current metric $m_k$ is that fewer points belonging to different spheres in a vertex in the graph indicates a preservation of local as well as global topology. With a perfect separation of manifolds, every node will only contain points belonging to one (in our case, one sphere). 
\section{Results}

\begin{figure}[h]
\centerline{\includegraphics[width=60mm]{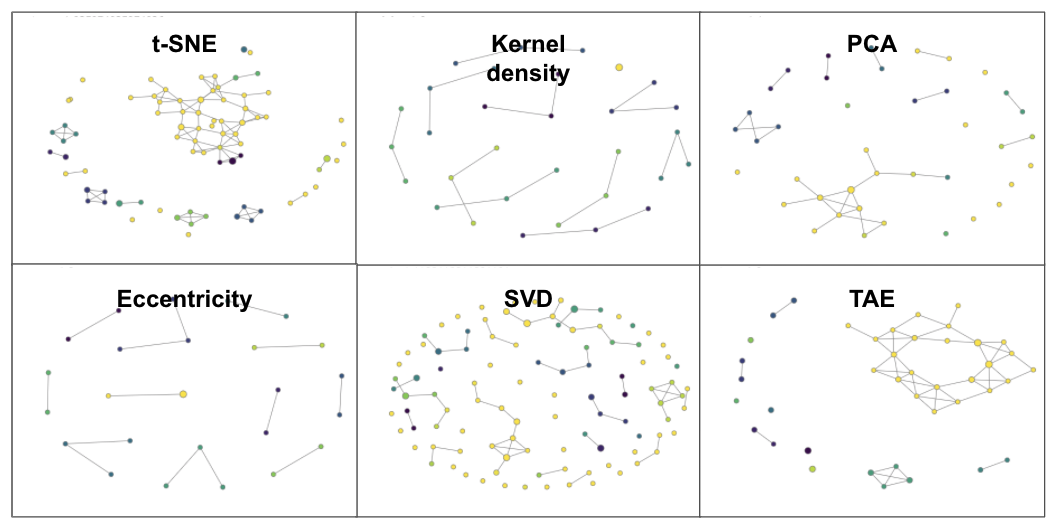}}
\caption{Examples of Mapper graphs for each filtration with the best results from table \ref{table:res}. Node color indicate average label value of which sphere points in the node belong to - yellow is the large sphere, others are the smaller. }
\label{fig:mappers}
\end{figure}

Table \ref{table:res} display our initial results, indicating that TAEs are better at allocating the different spheres into separate parts than the other methods, capturing more of the topological properties of the data. Although t-SNE is close, it does not completely separate the spheres like the TAE. In figure \ref{fig:mappers} we provide visualizations of the best results from table \ref{table:res}, where we can see that the TAE forms a graph in which somewhat of a circle can be distinguished for the larger sphere, and simultaneously separate the smaller spheres to become separate. The kernel density and the eccentricity also achieves optimal results according to the metric, but for these optimal results, they do not capture the inherent shape as well as the TAE (bottom right).

\begin{table}[]
\centering
\caption{Best average number of spheres found in each node. Lower is better, 1 is optimal. }
\label{table:res}
\begin{tabular}{|l|l|l|l|l|l|l|}
\hline
\textbf{}        & \textbf{TAE}   & \textbf{PCA} & \textbf{T-SNE} & \textbf{SVD} & \textbf{Kernel Density} & \textbf{Eccentricity} \\ \hline
\textbf{Best}    & \textbf{1.000} & 1.1          & 1.026          & 1.119        & \textbf{1.0}            & \textbf{1.0}          \\ \hline
\textbf{Average} & \textbf{1.021} & 1.206        & 1.130          & 1.205        & 1.621                   & 1.137                 \\ \hline
\textbf{Worst}   & \textbf{1.115} & 1.311        & 1.259          & 1.311        & 2.25                    & 1.3                   \\ \hline
\end{tabular}
\end{table}

\begin{figure}[h]
\centerline{\includegraphics[width=100mm]{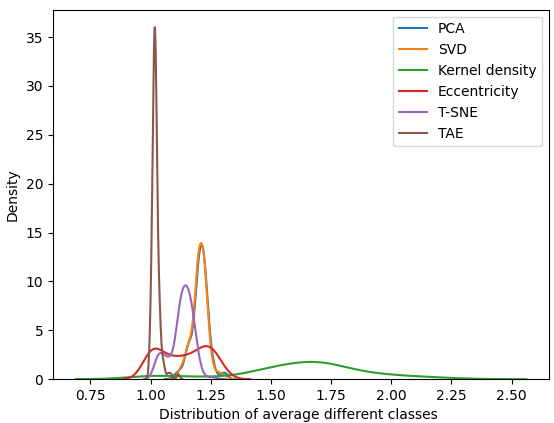}}
\caption{Distribution of the performance over the grid search for the different filtering functions. As can be seen, the TAE is significantly more centered towards the optimal value of the metric used.  }
\label{fig:dists}
\end{figure}

By looking at the distribution of the metric $m_k$ in figure \ref{fig:dists} for the different filter functions, we can see that the TAE has a distribution significantly closer to the optimal result (being 1) compared to the other filtering functions. This shows that over a large range of parameters, the TAE outperforms the rest in generating representations separating the different spheres into different nodes.

\section{Discussion \& future work}

In this extended abstract, we have shown that topological autencoders are superior to other filtering functions in terms of representing high-dimensional data when combined with the Mapper algorithm on a set of manifolds. We plan to further extend the experiments to more simulated manifold datasets as well as the other datasets applied in \cite{moor2019topological} to confirm whether this holds generally. Furthermore, the metrics used to evaluate how a filtering function best captures the topology of the data as well as its robustness will be refined, which could elaborate the way to benchmark different filtering functions and their ability to better preserve the topology of higher-dimensional data. In essence, it will be beneficial to aim for a method preserving global and local topology while simultaneously gaining an empirical understanding of different filtering functions' strengths and shortcomings. 

\section*{Acknowledgements}

This work was partially supported by Wallenberg Autonomous Systems Program (WASP). 

\medskip
\bibliography{neuripsref}

\begin{thebibliography}{9}
\providecommand{\natexlab}[1]{#1}
\providecommand{\url}[1]{\texttt{#1}}
\expandafter\ifx\csname urlstyle\endcsname\relax
  \providecommand{\doi}[1]{doi: #1}\else
  \providecommand{\doi}{doi: \begingroup \urlstyle{rm}\Url}\fi

\bibitem[Singh et~al.(2007)Singh, M{\'e}moli, and
  Carlsson]{singh2007topological}
Gurjeet Singh, Facundo M{\'e}moli, and Gunnar~E Carlsson.
\newblock Topological methods for the analysis of high dimensional data sets
  and 3d object recognition.
\newblock \emph{SPBG}, 91:\penalty0 100, 2007.

\bibitem[Munch and Wang(2015)]{munch2015convergence}
Elizabeth Munch and Bei Wang.
\newblock Convergence between categorical representations of reeb space and
  mapper.
\newblock \emph{arXiv preprint arXiv:1512.04108}, 2015.

\bibitem[Kumari et~al.(2020)Kumari, Rupela, Gupta, and
  Krishnamurthy]{kumari2020shapevis}
Nupur Kumari, Akash Rupela, Piyush Gupta, and Balaji Krishnamurthy.
\newblock Shapevis: High-dimensional data visualization at scale.
\newblock In \emph{Proceedings of The Web Conference 2020}, pages 2920--2926,
  2020.

\bibitem[Carriere et~al.(2018)Carriere, Michel, and
  Oudot]{carriere2018statistical}
Mathieu Carriere, Bertrand Michel, and Steve Oudot.
\newblock Statistical analysis and parameter selection for mapper.
\newblock \emph{The Journal of Machine Learning Research}, 19\penalty0
  (1):\penalty0 478--516, 2018.

\bibitem[Lum et~al.(2013)Lum, Singh, Lehman, Ishkanov, Vejdemo-Johansson,
  Alagappan, Carlsson, and
  Carlsson]{lum_singh_lehman_ishkanov_vejdemo-johansson_alagappan_carlsson_carlsson_2013}
P.~Y. Lum, G.~Singh, A.~Lehman, T.~Ishkanov, M.~Vejdemo-Johansson,
  M.~Alagappan, J.~Carlsson, and G.~Carlsson.
\newblock Extracting insights from the shape of complex data using topology.
\newblock \emph{Scientific Reports}, 3\penalty0 (1), 2013.
\newblock ISSN 2045-2322.
\newblock \doi{10.1038/srep01236}.
\newblock URL \url{https://dx.doi.org/10.1038/srep01236}.

\bibitem[Maaten and Hinton(2008)]{maaten2008visualizing}
Laurens van~der Maaten and Geoffrey Hinton.
\newblock Visualizing data using t-sne.
\newblock \emph{Journal of machine learning research}, 9\penalty0
  (Nov):\penalty0 2579--2605, 2008.

\bibitem[Tenenbaum et~al.(2000)Tenenbaum, De~Silva, and
  Langford]{tenenbaum2000global}
Joshua~B Tenenbaum, Vin De~Silva, and John~C Langford.
\newblock A global geometric framework for nonlinear dimensionality reduction.
\newblock \emph{science}, 290\penalty0 (5500):\penalty0 2319--2323, 2000.

\bibitem[Moor et~al.(2019)Moor, Horn, Rieck, and
  Borgwardt]{moor2019topological}
Michael Moor, Max Horn, Bastian Rieck, and Karsten Borgwardt.
\newblock Topological autoencoders.
\newblock \emph{arXiv preprint arXiv:1906.00722}, 2019.

\bibitem[Ram et~al.(2010)Ram, Jalal, Jalal, and Kumar]{ram2010density}
Anant Ram, Sunita Jalal, Anand~S Jalal, and Manoj Kumar.
\newblock A density based algorithm for discovering density varied clusters in
  large spatial databases.
\newblock \emph{International Journal of Computer Applications}, 3\penalty0
  (6):\penalty0 1--4, 2010.

\end{thebibliography}



\end{document}